\documentclass[aps,twocolumn,showpacs,preprintnumbers,nofootinbib,prd,10pt,superscriptaddress,floats,floatfix]{revtex4-2}

\makeatletter
\def\l@subsubsection#1#2{}
\def\l@subsubsubsection#1#2{}
\makeatother
\usepackage{graphicx,amssymb,amsmath,amsthm,amsfonts,epsfig,epsf,fixmath,cancel}
\usepackage{comment}
\usepackage{mathrsfs}
\usepackage[usenames]{color}
\usepackage{epstopdf}
\usepackage{aas_macros}
\usepackage{bm}
\usepackage{dcolumn}
\usepackage{etoolbox}
\usepackage{latexsym}
\usepackage{orcidlink}
\usepackage{tabularx}
\usepackage{booktabs}
\usepackage{longtable}
\def\arraystretch{1.25}
\usepackage{enumerate}
\usepackage{tensor,multirow}
\usepackage{url}
\usepackage{float}
\usepackage{xcolor}
\usepackage{nicefrac}
\definecolor{MidnightBlue}{RGB}{25, 25, 112}
\hypersetup{colorlinks=true, citecolor=MidnightBlue,
            linkcolor=MidnightBlue, urlcolor=MidnightBlue, linktocpage=true}

\usepackage{silence}
\newcommand{\be}{\begin{equation}}
\newcommand{\eeq}{\end{equation}}
\newcommand{\ba}{\begin{align}}
\newcommand{\ea}{\end{align}}

\def\BH{\text{\tiny BH}}

\newcommand{\GSSI}{Gran Sasso Science Institute (GSSI), I-67100 L’Aquila, Italy}
\newcommand{\GranSasso}{INFN, Laboratori Nazionali del Gran Sasso, I-67100 Assergi, Italy}
\newcommand{\jhu}{William H.\ Miller III Department of Physics and Astronomy, Johns Hopkins University, \\ 3400 N. Charles Street, Baltimore, Maryland, 21218, USA}
\newcommand{\UPenn}{Center for Particle Cosmology, Department of Physics and Astronomy, University of Pennsylvania 209 South 33rd Street, Philadelphia, Pennsylvania 19104, USA}

\begin{document}
\title{Systematic biases from ignoring environmental tidal effects\texorpdfstring{\\}{ }in gravitational wave observations}

\author{Valerio De Luca\hspace{0.05cm}\orcidlink{0000-0002-1444-5372}}
\email{vdeluca@sas.upenn.edu}
\affiliation{\UPenn}

\author{Loris Del Grosso\hspace{0.05cm}\orcidlink{0000-0002-6722-4629}}
\email{ldelgro1@jh.edu}
\affiliation{\jhu}

\author{Francesco Iacovelli\hspace{0.05cm}\orcidlink{0000-0002-4875-5862}}
\email{fiacovelli@jhu.edu}
\affiliation{\jhu}

\author{Andrea Maselli\hspace{0.05cm}\orcidlink{0000-0001-8515-8525}}
\email{andrea.maselli@gssi.it}
\affiliation{\GSSI}
\affiliation{\GranSasso}

\author{Emanuele Berti\hspace{0.05cm}\orcidlink{0000-0003-0751-5130}}
\email{berti@jhu.edu}
\affiliation{\jhu}

\begin{abstract}
\noindent
Binary black hole systems are typically assumed to evolve in vacuum. However, the environment surrounding the binary components can influence their properties, such as their tidal deformability, affecting the gravitational waveform produced by the binary and its interpretation in gravitational wave data analysis.
In this work we focus on next-generation experiments, such as the Einstein Telescope and LISA, and we quantify the systematic biases in gravitational wave observations that arise when tidally deformed binaries are interpreted as occurring in vacuum. We consider binaries over a range of masses and we compare different phenomenological models for the dynamical evolution of the tidal deformability. We find that systematic biases could significantly affect the measurability of the binary parameters if tidal effects are not carefully modeled.
\end{abstract}

\preprint{ET-0067A-25}
\maketitle

\section{Introduction}

Gravitational wave (GW) astronomy is profoundly advancing our understanding of general relativity and astrophysics by enabling the direct detection of compact object mergers. The LIGO-Virgo-KAGRA detector network has already recorded nearly 100 binary black hole (BH) mergers~\cite{LIGOScientific:2018mvr,LIGOScientific:2020ibl, KAGRA:2021vkt}, and this number is expected to grow dramatically with the advent of next-generation GW detectors, including the Einstein Telescope (ET), Cosmic Explorer (CE), and LISA~\cite{Punturo:2010zz, Hild:2010id, Essick:2017wyl, Sathyaprakash:2019yqt, Maggiore:2019uih, Reitze:2019iox, Kalogera:2021bya, Evans:2021gyd, Evans:2023euw, Branchesi:2023mws,LISA:2017pwj,LISA:2024hlh}.

The unprecedented science accessible through these experiments relies upon accurate measurements of the binary parameters, which, in turn, depend on the waveform models used to analyze the GW signals. Mismatched waveforms can potentially cause wrong interpretations of binary events and introduce biases in parameter estimation~\cite{Cutler:2007mi}.
For instance, imperfect waveform models, based on semi-analytic approximations of two-body solutions in general relativity calibrated to numerical relativity simulations, can lead to systematic biases in the inferred binary parameters. Statistical errors decrease with signal-to-noise ratio (SNR) while systematic errors do not, and therefore this issue will become more pressing as the sensitivity of GW detectors increases~\cite{Owen:2023mid, Read:2023hkv, Purrer:2019jcp, Berti:2006ew, Kapil:2024zdn, Dhani:2024jja}.

In this work we discuss the relevance of environmental effects in the interpretation of GW events and the mismatch with waveform models assuming that the binary evolves in vacuum, i.e., in the absence of an external environment. The effect of environments on GW signals driven by secular phenomena such as baryonic accretion, superradiance, and dark matter physics has been studied in Refs.~\cite{Barausse:2014pra, Barausse:2014tra, Toubiana:2020drf, Cole:2022yzw, CanevaSantoro:2023aol, Khalvati:2024tzz, Zwick:2022dih, Roy:2024rhe, Garg:2024qxq}. In this paper, we mostly focus on the systematic biases induced by ignoring environmental tidal effects in gravitational waveforms. 

Tidal effects are widely recognized as a crucial factor in the inspiral of binary systems. Their leading contribution to the gravitational waveform appears at fifth post-Newtonian (PN) order~\cite{Flanagan:2007ix,Hinderer:2009ca}, and their measurement provides valuable insight into the internal structure of compact objects. For example, tidal effects have been used to probe the equation of state of neutron stars~\cite{GuerraChaves:2019foa, Chatziioannou:2020pqz}, and they could reveal the existence of exotic compact objects or new physics~\cite{Cardoso:2017cfl, Maselli:2017cmm, Cardoso:2019rvt, Herdeiro:2020kba, Chen:2023vet, Berti:2024moe, Maselli:2018fay, Datta:2021hvm, Suarez-Fontanella:2024epb}. Within the realm of linear response theory, tidal effects are usually parameterized in terms of a set of coefficients called tidal Love numbers (TLNs)~\cite{1909MNRAS..69..476L, Hinderer:2007mb, Binnington:2009bb, Damour:2009vw} which, for various families of asymptotically flat BHs in four-dimensional space-times, are found to vanish exactly~\cite{Deruelle:1984hq,Binnington:2009bb, Damour:2009vw, Damour:2009va, Pani:2015hfa, Pani:2015nua, Gurlebeck:2015xpa, Porto:2016zng, LeTiec:2020spy, Chia:2020yla, LeTiec:2020bos, Hui:2020xxx, Charalambous:2021mea, Charalambous:2021kcz, Creci:2021rkz, Bonelli:2021uvf, Ivanov:2022hlo, Charalambous:2022rre, Katagiri:2022vyz, Ivanov:2022qqt, Berens:2022ebl, Bhatt:2023zsy, Sharma:2024hlz, Rai:2024lho}. This property has also been studied at the nonlinear level~\cite{DeLuca:2023mio, Riva:2023rcm, Iteanu:2024dvx}, with results demonstrating that it remains valid at every perturbative order in general relativity~\cite{Kehagias:2024rtz, Combaluzier-Szteinsznaider:2024sgb, Gounis:2024hcm}. The vanishing of the TLNs can, however, be violated in scenarios involving BH mimickers~\cite{Pani:2015tga, Cardoso:2017cfl, Nair:2022xfm, Avitan:2023txy, DeLuca:2024nih, Coviello:2025pla}, in the presence of a cosmological constant, in theories of modified gravity, and in higher dimensions~\cite{Nair:2024mya, Cardoso:2017cfl, Cardoso:2018ptl, DeLuca:2022tkm, Barura:2024uog, Kol:2011vg, Cardoso:2019vof, Hui:2020xxx, Rodriguez:2023xjd, Charalambous:2023jgq, Charalambous:2024tdj, Charalambous:2024gpf, Ma:2024few, Barbosa:2025uau}. The nature of TLNs has also been investigated in the presence of  time-dependent tidal fields~\cite{Saketh:2023bul, Perry:2023wmm, Chakraborty:2023zed, Ivanov:2024sds, DeLuca:2024ufn, Bhatt:2024yyz, Katagiri:2024wbg, Katagiri:2024fpn, Bhatt:2024rpx}. 

\begin{figure*}[t] 
    \includegraphics[width=0.8\textwidth]{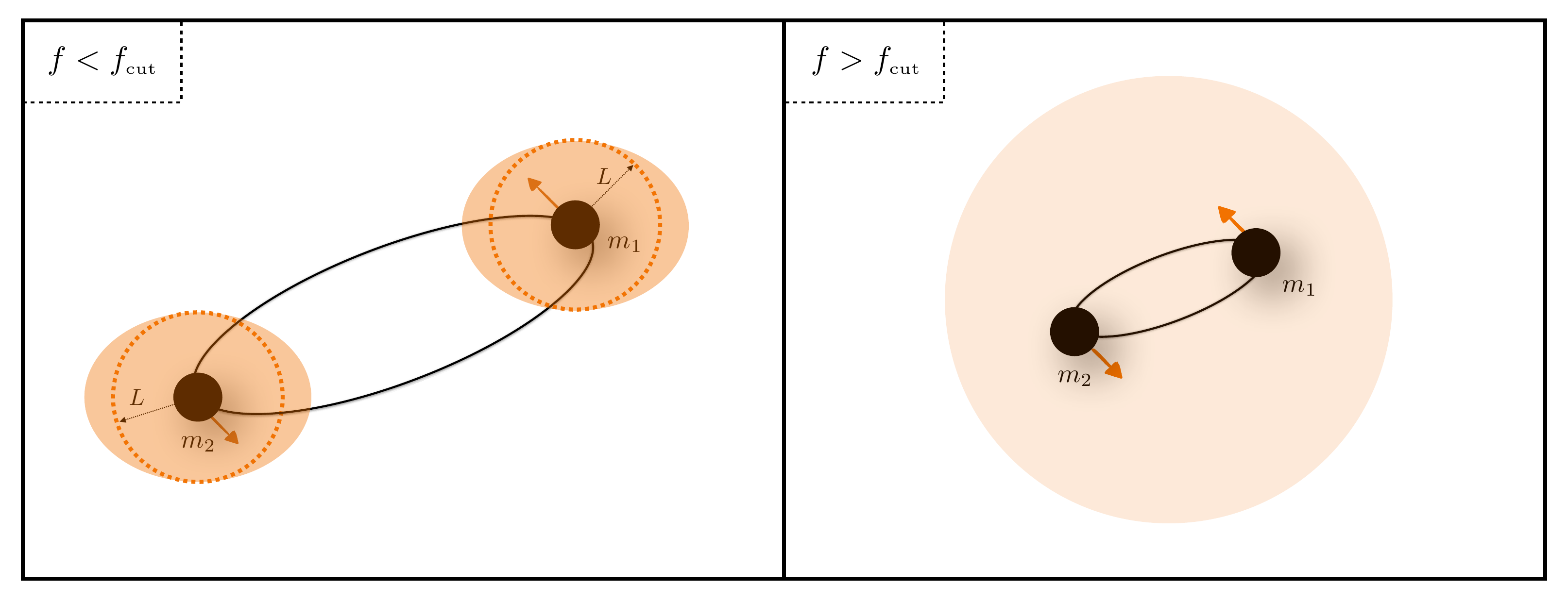}
    \caption{Schematic illustration of the binary system under consideration, consisting of two BHs (black circles) embedded in external environments (orange halos). In the left panel, prior to reaching the Roche frequency $f_\text{\tiny cut}$, the binary undergoes tidal interactions characterized by a nonzero tidal Love number. After tidal disruption (right panel), the BHs evolve as isolated (``naked'') objects within a significantly more diffuse halo. Further details are provided in Appendix~\ref{app-A}.}
\label{fig:setup}
\end{figure*}

The presence of an external environment has been shown to induce a nonzero TLN in nonvacuum binary BHs~\cite{Baumann:2018vus, DeLuca:2021ite, DeLuca:2022xlz, Brito:2023pyl, Capuano:2024qhv,Cardoso:2019upw, Cardoso:2021wlq, Katagiri:2023yzm, DeLuca:2024uju, Cannizzaro:2024fpz}. This effect may become significant for dense environments, before their possible disruption due to tidal interactions, and leave a distinctive imprint on the gravitational waveform during the inspiral phase. If omitted in GW data analysis, it can result in misidentification of the features of the coalescing compact objects, and introduce substantial biases in their parameters. The goal of this work is to quantify the systematic biases arising from the omission of environmental tidal effects.

The manuscript is organized as follows. In Sec.~\ref{sec:env} we provide the general framework to describe environmental effects in gravitational waveforms and outline the data analysis formalism. In Sec.~\ref{Sec:num} we compare waveforms in vacuum and in different environments for binaries detectable by ET and by LISA. Section~\ref{sec:conc} is devoted to the conclusions. Some technical details are discussed in three Appendixes. Unless specified otherwise, we adopt geometrical units ($G = c = 1$).

\section{Environmental effects: setup and data analysis}\label{sec:env}

In this work, we investigate the dynamics of equal-mass binary BH systems surrounded by small-scale, localized environments, initially assumed to be in stationary equilibrium around each BH. Such environments are astrophysically motivated and may arise in a variety of contexts, including accretion disks, superradiant clouds, or dense dark matter substructures bound to individual BHs. This setup is physically distinct from scenarios involving circumbinary structures -- such as those studied in Ref.~\cite{DOrazio:2021kob} -- where the binary evolves within a common disk from the outset~\cite{Thun_2017}: for systems evolving in circumbinary structures, the effects of the environment are not limited to the ones considered in this paper, but they include also accretion, dynamical friction and drag forces, even during the early stages of the inspiral.

We consider equal-mass binary BH systems with component masses $m_1 = m_2 = m_{\text{\tiny{\rm BH}}}$. Each BH is embedded in an external environment of mass $m_{\rm env}$, characterized by a fundamental length scale $L$ corresponding, for instance, to the location of maximum density of the matter distribution. We assume the mass of the environment to be small compared to $m_\text{\tiny BH}$, i.e., we consider physical scenarios in which the parameter $\epsilon \equiv m_{\rm env} / m_{\text{\tiny{\rm BH}}} \ll 1$, and neglect terms of order $\mathcal{O}(\epsilon^2)$.
A schematic picture of the setup under investigation is sketched in Fig.~\ref{fig:setup}.

The presence of the environment induces a nonvanishing tidal deformability 
on each BH~\cite{Baumann:2018vus, DeLuca:2021ite, DeLuca:2022xlz, Brito:2023pyl, Cannizzaro:2024fpz, Cardoso:2019upw, Cardoso:2021wlq, Katagiri:2023yzm, DeLuca:2024uju}, leaving an imprint on the emitted GW signal. Tidal contributions 
enter the waveform at the $5\,\rm{PN}$ and $6\,\rm{PN}$ orders~\cite{Damour:1986ny, Flanagan:2007ix, Vines:2011ud} through the following terms:
\begin{equation}\label{eq:5PN}
    \begin{aligned}
        \psi_{\rm{tidal}} = -\dfrac{3}{128 \eta x^{\nicefrac{5}{2}}} \bigg[ \frac{39}{2} \tilde{\Lambda}  x^5 &+ \frac{3115}{64} \tilde{\Lambda}x^6\\
        &-\frac{6595}{364} \delta \tilde{\Lambda}\sqrt{1-4 \eta } x^6 \bigg]  \,,
    \end{aligned}
\end{equation}
where $f$ is the GW frequency, $x = (\pi M f)^{\nicefrac{2}{3}}$, and $M = \mathcal{M} \eta^{\nicefrac{-3}{5}} = 2 m_{\text{\tiny{\rm BH}}}(1+\epsilon)$ is the total mass of the binary. The chirp mass is defined by $\mathcal{M} = (m_1 m_2)^{\nicefrac{3}{5}}/(m_1+m_2)^{\nicefrac{1}{5}}$, and the symmetric mass ratio by $\eta = m_1 m_2 / (m_1+m_2)^2$. 
For equal-mass binaries (with $\mathcal{M}=m_{\text{\tiny{\rm BH}}} / 2^{\nicefrac{1}{5}}$ and $\eta=1/4$) the last contribution in Eq.~\eqref{eq:5PN} vanishes, 
and $\psi_{\rm tidal}$ is fully determined by the tidal deformability 
$\tilde{\Lambda}$, which  depends on the dimensionless individual quadrupolar TLNs of the two BHs,  $k_{2,i}$ ($i=1,2$)~\cite{Flanagan:2007ix,Vines:2011ud, Wade:2014vqa}, normalized by their masses~\cite{Cardoso:2017cfl}, and is given by:
\begin{equation}
    \begin{aligned}
        \tilde{\Lambda} &=  \frac{8}{13}\left[\left(1+7\eta-31\eta^2\right) \frac{2}{3} \left(k_{2,1}+k_{2,2}\right) \right . \\
        &\left . \quad\;+\sqrt{1-4\eta}\left(1+9\eta-11\eta^2\right) \frac{2}{3} \left(k_{2,1}-k_{2,2}\right) \right]\,.
    \end{aligned}
\label{eq:lambdatilde}
\end{equation}
Assuming that the environments around the two BHs have identical properties, we set $k_{2,1} = k_{2,2} = k_2$, which simplifies Eq.~\eqref{eq:lambdatilde} for equal-mass systems to 
$\tilde{\Lambda} = 2 k_2 / 3$. 

The TLN $k_2$ generally depends on the type of environment. 
However, as suggested in Refs.~\cite{Mora:2003wt}, in the limit where the 
characteristic scale $L$ is much larger than the BH radius $r_{\text{\tiny{\rm BH}}} = 2\, m_{\text{\tiny{\rm BH}}}$, TLNs scale as $k_2 \propto (L/m_{\text{\tiny{\rm BH}}})^5$, independent of the specific 
matter distribution (see also Refs.~\cite{Baumann:2018vus, DeLuca:2021ite, DeLuca:2022xlz, Brito:2023pyl, Cannizzaro:2024fpz, Cardoso:2019upw, Cardoso:2021wlq, Katagiri:2023yzm, DeLuca:2024uju} for similar analyses). Moreover, $k_2$ must vanish in the vacuum limit $\epsilon\to 0$, as expected for BHs in GR~\cite{Binnington:2009bb,Damour:2009vw,Damour:2009va,Pani:2015hfa,Pani:2015nua,Gurlebeck:2015xpa,Porto:2016zng,LeTiec:2020spy, Chia:2020yla,LeTiec:2020bos,Charalambous:2021mea,Charalambous:2021kcz,Ivanov:2022hlo,Charalambous:2022rre,Katagiri:2022vyz, Ivanov:2022qqt,Berens:2022ebl, Bhatt:2023zsy, Sharma:2024hlz}.

We exploit these properties and make our analysis fully 
agnostic by parameterizing Eq.~\eqref{eq:lambdatilde} as:
\begin{equation}
    \tilde{\Lambda}(\epsilon,L)=\frac{2}{3}{\cal F}\epsilon \, \tilde{L}^5\, ,\label{eq:tidalparamet}
\end{equation}
where $\tilde{L} = L/m_{\text{\tiny{\rm BH}}}$. 
The numerical coefficient $\mathcal{F}$, which is independent of $\epsilon$ and $L$, 
can be mapped to specific models of the environment. For example, for a thin shell of pressureless dust 
surrounding a Schwarzschild BH, $\mathcal{F} = 1$~\cite{Cardoso:2019upw}, while for a 
gravitational atom in the lowest energy mode, $\mathcal{F} = 15/2$~\cite{Arana:2024kaz}. Hereafter, we rescale Eq.~\eqref{eq:tidalparamet} such that 
$\epsilon\rightarrow{\cal F}\epsilon$, which is equivalent to setting ${\cal F}=1$.

In realistic scenarios, $\tilde{L} \gg 1$, since the environment can remain in equilibrium bounded with a BH only if it is located beyond the innermost stable circular orbit (ISCO). For a nonspinning BH, this condition implies $\tilde{L} \gtrsim 6$. Notice that BHs with small-sized environments would have TLNs potentially smaller than 
those of neutron stars with realistic equation of states~\cite{Hinderer:2009ca,GuerraChaves:2019foa, Chatziioannou:2020pqz} or non-ultracompact exotic compact objects~\cite{Cardoso:2017cfl, Cardoso:2019rvt, Herdeiro:2020kba, Chen:2023vet, Berti:2024moe}, while in principle large values of $\tilde{L}$ can produce a sizable tidal deformability in the dressed BH systems.

We account for the possible tidal disruption of the environment during the binary evolution, which affects both the tidal deformability and the mass of the matter ``dressing'' the BH.  To model this effect, we follow Ref.~\cite{DeLuca:2022xlz} and introduce the frequency-dependent parameters
\begin{equation}
    \tilde{\Lambda} \to \tilde{\Lambda}\, {\cal S}(f) = \tilde{\Lambda}\,\left[ \frac{1 + e^{\nicefrac{-f_\text{\tiny cut}}{f_\text{\tiny slope}}}}{1 + e^{\nicefrac{(f - f_\text{\tiny cut})}{f_\text{\tiny slope}}}} \right]\,, \label{tapering}
\end{equation}
and
\begin{equation}
\label{smoothingM}
   m_{\text{\tiny{\rm BH}}}\!\to m_{\text{\tiny{\rm BH}}} \, \mathcal{S}_m (f) \! = m_{\text{\tiny{\rm BH}}} \!\left[\!\frac{(1+ \epsilon)\!+e^{\nicefrac{(f-f_\text{\tiny cut})}{f_\text{\tiny slope}}}}{1+e^{\nicefrac{(f-f_\text{\tiny cut})}{f_\text{\tiny slope}}}}\!\right]\, .
\end{equation}
The smoothing functions ${\cal S} (f)$ and ${\cal S}_{m} (f)$ ensure that $\tilde{\Lambda}$ 
vanishes smoothly around the Roche frequency $f_\text{\tiny cut}$, while the BH mass reduces 
to its vacuum value.\footnote{To improve numerical stability in the \texttt{Python} implementation of the code, we adopt a polynomial filter that mimics the sigmoid behavior of Eqs.~\eqref{tapering} and~\eqref{smoothingM}, while being easily differentiable and exactly vanishing at the edges. We verified that this does not introduce any difference in the results.} The rate at which this transition 
occurs is controlled by the characteristic slope $f_\text{\tiny slope}$. We set 
$f_\text{\tiny slope} =f_\text{\tiny cut}/5$, but we have verified that its specific value does not 
significantly affect our analysis (see also Ref.~\cite{DeLuca:2022xlz}). For an equal-mass binary, $f_\text{\tiny cut}$ can be 
expressed as~\cite{Cannizzaro:2024fpz}
\begin{equation}
\label{fcut}
    f_\text{\tiny cut} = \frac{1}{\pi} \sqrt{\frac{2 m_{\text{\tiny{\rm BH}}}(1+\epsilon)}{d_{\text{\tiny{\rm Roche}}}^{\,3}}}
    = \sqrt{\frac{2}{\gamma^3}} \frac{1}{\pi \,m_{\text{\tiny{\rm BH}}}} \frac{(1+\epsilon)^{\nicefrac{1}{2}}}{\tilde{L}^{\nicefrac{3}{2}}}\,,
\end{equation}
where $d_{\text{\tiny{\rm Roche}}} = \gamma \, m_{\text{\tiny{\rm BH}}} \tilde{L}$ is the Roche 
radius of each dressed BH. The numerical parameter $\gamma$ takes values ranging from 1.26 for 
rigid bodies to 2.44 for fluid ones~\cite{Shapiro:1983du}. We will assume $\gamma = 2.44$, as it is more accurate in realistic scenarios. For comparison, we also give the GW frequency at the ISCO of the binary
\begin{equation}\label{eq:ISCO_freq}
    f_\text{\tiny ISCO} \simeq 4.4 \, {\rm kHz} \, {\rm M_\odot} / (m_1 + m_2)\,,
\end{equation}
which is always bigger than $f_\text{\tiny cut}$ in our scenario. In Appendix~\ref{app-A} we discuss the interaction of the binary system with the tidally destroyed environment following the fading process, showing that GW emission predominantly drives the system's dynamics for the parameter range considered in this study. This is due to the fact that, by the time the environment has been largely depleted and a circumbinary structure has formed, the binary’s orbital frequency has increased sufficiently, rendering environmental effects -- such as accretion or dynamical friction -- negligible.

As we will show below, the smoothing of the masses and of the tidal deformability can significantly impact the GW signal and the inference of the source parameters.

\subsection{Waveform model and parameter estimation}

We model the GW signal emitted during the coalescence through the \textsc{IMRPhenomD} waveform family~\cite{Husa:2015iqa,Khan:2015jqa}, which takes into account the inspiral, merger and ringdown phases of the coalescence, augmented by the inclusion of frequency-dependent environmental effects in the tidal GW phase $\psi_{\rm tidal} (f)$ and in the BH masses~\cite{DeLuca:2022xlz}.
In Fourier space, the template reads~\cite{Sathyaprakash:1991mt, Damour:2000gg}
\begin{equation}\label{eq:htilde}
    \tilde h (f;\bm{\theta}) = C_\Omega(f;\bm{\theta}) {\cal A} (f;\bm{\theta}) \, e^{{\rm i} \psi_\text{\tiny PP} (f;\bm{\theta}) + {\rm i} \psi_{\rm tidal} (f;\bm{\theta})}\,,
\end{equation}
where $\bm{\theta}$ is the set of intrinsic and extrinsic source parameters. The amplitude ${\cal A} (f;\bm{\theta})$ and point-particle phase $\psi_\text{\tiny PP}$ are the outputs of the chosen spin-aligned waveform, which also depends on the individual spin components aligned with the orbital angular momentum of the binary, $\chi_{1,z}$ and $\chi_{2,z}$.
In our analysis we will assume nonspinning binaries ($\chi_{1,z} = \chi_{2,z} = 0$), although the aligned spin components are included as waveform parameters.

The leading term in the waveform amplitude is given by~\cite{Berti:2004bd}
\begin{equation}
    \mathcal{A} (f) = \sqrt{\frac{5}{24}} \frac{\mathcal{M}^{\nicefrac{5}{6}} \mathcal{S}_m^{\nicefrac{5}{6}} (f)}{\pi^{\nicefrac{2}{3}}d_L f^{\nicefrac{7}{6}}} \,,
\end{equation}
where $d_L$ is the luminosity distance, and the smoothing factor ${\cal S}_m$ takes into account changes in the chirp mass due to the fading environment. 
The geometric coefficient for the fundamental mode $C_{\Omega}=[F_+^2 (1+\cos^2 \iota)^2 + 4 F_\times^2 \cos \iota]^{\nicefrac{1}{2}}$ in Eq.~\eqref{eq:htilde} depends on the inclination angle $\iota$ between the binary's line of sight and its orbital angular momentum, as well as on the detector's antenna pattern functions $F_{+,\times} (f;\vartheta, \varphi, \psi)$, which are functions of the position of the source in the sky $(\vartheta,\, \varphi)$ and of the polarization angle $\psi$, as well as of time, and thus frequency (see e.g. the discussion in Sec.~3.2 of~\cite{Iacovelli:2022bbs}). 
Their explicit expressions depend on the GW detector under consideration: for ET we use the expression given e.g. in Ref.~\cite{Iacovelli:2022bbs}, also including the modulations due to the Earth's rotation, while for LISA we work in the so-called $AET$ basis~\cite{Prince:2002hp}, with the same expressions detailed in Ref.~\cite{Marsat:2020rtl}. Both are implemented in the \texttt{gwfast} package~\cite{Iacovelli:2022mbg}.

To predict the measurement accuracy of the source parameters, we use the Fisher information matrix approach~\cite{Cutler:1994ys, Poisson:1995ef, Vallisneri:2007ev}. In this framework, for GW signals with a high SNR, the posterior distribution of the model parameters approximates a multivariate Gaussian centered at their true values, $\bm{\hat{\theta}}$, with a covariance matrix given by ${\Sigma}_{ij} = (\Gamma_{ij})^{-1}$, where
\begin{equation}
\label{eq:fisher_def}
  \Gamma_{ij}= \left\langle \frac{\partial h}{\partial \theta_i}\bigg\vert\frac{\partial h}{\partial \theta_j}\right\rangle_{\bm{\theta}=\bm{\hat{\theta}}}
\end{equation}
is the Fisher matrix. The statistical uncertainty of the $i$\textsuperscript{th} parameter is then $\sigma_i = \Sigma^{\nicefrac{1}{2}}_{ii}$. The Fisher matrix is defined via the scalar product of two waveform templates $h_1$ and $h_2$, weighted by the detector noise power spectral density (PSD) $S_n(f)$ under the assumption of stationary, Gaussian noise:
\begin{equation}
    \langle h_1\vert h_2\rangle = 4\Re\int_{f_\text{\tiny min}}^{f_\text{\tiny max}} \frac{\tilde{h}_1^\star(f)\tilde{h}_2(f)}{S_n(f)} {\rm d}f \,, \label{scalprod}
\end{equation}
where a $\star$ denotes complex conjugation. The SNR for a given set of binary parameters is then defined as $\rho = \langle h \vert h \rangle^{\nicefrac{1}{2}}$. The integration limits $f_\text{\tiny min}$ and $f_\text{\tiny max}$ in Eq.~\eqref{scalprod}  depend on the specific GW experiment. For ET and LISA, we use $f_\text{\tiny min} = 2\,{\rm Hz}$ and $f_\text{\tiny min} = 10^{-5}\,{\rm Hz}$, respectively, while for both cases we set $f_\text{\tiny max}$ to a frequency well after the merger if the signal lies entirely in the detector band, or to $f_\text{\tiny max}= 4096\,{\rm Hz}$ and $f_\text{\tiny max}= 0.3\,{\rm Hz}$, respectively. For LISA, we further limit the observational time to 4~yr. As for the noise PSDs, for ET we use the same sensitivity adopted in Ref.~\cite{Branchesi:2023mws} for a $10\,{\rm km}$ detector with both a high-frequency and a low-frequency instrument, and consider a triangular geometry. 
We have verified that changing detector configuration, i.e. assuming two L-shaped instruments~\cite{Branchesi:2023mws}, or adopting CE as the next-generation observatory, does not significantly affect our results.
For LISA, we use the same analytical prescription for the sensitivity provided in Ref.~\cite{Babak:2021mhe}, further including the contribution of the Galactic White Dwarf confusion noise, and considering $2.5\times10^6\,{\rm km}$ equal arms. 

The fading of the environment encoded in Eqs.~\eqref{tapering} and~\eqref{smoothingM} imposes a hierarchy of frequencies, such that $f_\text{\tiny min} < f_\text{\tiny cut} < f_\text{\tiny max}$. By imposing $f_\text{\tiny min} < f_\text{\tiny cut}$ and neglecting $\mathcal{O}(\epsilon)$ terms, we get
\begin{equation}
\begin{aligned}
\tilde{L} \,\leq\, &\dfrac{1}{\gamma}\left[ \dfrac{2}{(\pi\, m_{\text{\tiny{\rm BH}}}\, f_\text{\tiny min})^2}\right]^{\nicefrac{1}{3}} \\ 
&\approx \left\{\begin{aligned}
      &50 \, \left( \dfrac{m_\text{\tiny BH}}{30\, {\rm M_\odot}} \right)^{\nicefrac{-2}{3}} \, \text{for ET}\,, \\
      &2\times10^2 \, \left( \dfrac{m_\text{\tiny BH}}{10^6\, {\rm M_\odot}} \right)^{\nicefrac{-2}{3}}\, \text{for LISA} \,.
  \end{aligned}\right.
\end{aligned}
\end{equation}
We will come back to this condition later in Sec.~\ref{Sec:num}. Similarly, as discussed above, the condition $f_\text{\tiny cut} < f_\text{\tiny max}$ is automatically satisfied upon having bounded environments in equilibrium before reaching the ISCO, i.e., $\tilde{L}\gtrsim 6$.

To summarize, our waveform model depends on thirteen binary parameters $\bm{\theta}=\{{\cal M},\,\eta,\,\chi_{1,z},\,\chi_{2,z},\,t_c,\,\phi_c,\,\vartheta,\, \varphi,\, \psi,\, \iota,\, d_L,\,\tilde \Lambda,\, f_\text{\tiny cut}\}$. Here, $(t_c,\,\phi_c)$ denote the coalescence time and phase, set to $t_c=\phi_c=0$ for definiteness. We also set $\psi = 0$, and choose $\vartheta,\, \varphi$, and $\iota$ to obtain the average response of the detector over the sky for each choice of the remaining parameters. The results for the spins are given in terms of the effective spin parameter $\chi_{\rm eff} = (m_1\chi_{1,z} + m_2\chi_{2,z})/(m_1 + m_2) \in [-1,\,1]$.

The Fisher matrix also allows us to estimate biases in the source parameters that may arise due to mismodeling of the recovery template relative to the real waveform present in the data. Such biases can also degrade detector sensitivity (see e.g. Ref.~\cite{Wang:2023qgw} for a study of highly spinning neutron star binaries, and Ref.~\cite{DeLuca:2024uju} for subsolar primordial BHs surrounded by thin matter shells), and their study is the main focus of this paper. 

{%
\renewcommand{\arraystretch}{1.4}
\setlength{\tabcolsep}{8.2pt}
\begin{table}[t]
    \begin{tabular}{c ll}
        \toprule
        \midrule
        Waveforms  & $h_{\rm rec}$ & $h_{\rm true}$ \\
        \midrule
        \midrule
        Case~1 & vacuum (no $\tilde{\Lambda}$) & vacuum + fixed $\tilde{\Lambda}$ \\
        Case~2 & vacuum (no $\tilde{\Lambda}$) & fading environment \\
        Case~3 & fixed environment & fading environment \\
        \midrule
        \bottomrule
    \end{tabular}
    \caption{Summary of the cases discussed in Sec.~\ref{Sec:num}. For each case, $h_{\rm true}$ and $h_{\rm rec}$ identify the \emph{true} signal present in the data, and the  \emph{recovery} template, respectively.}
\label{tab:res}
\end{table}
}%

In particular, we focus on the systematic errors introduced when using \emph{recovery} templates---either assuming vacuum or incorporating only partial knowledge of tidal effects---to interpret \emph{true} signals that encode environmental imprints. To compute these errors, we adopt the Cutler-Vallisneri approach~\cite{Cutler:2007mi}, defining\footnote{The expression~\eqref{bias} applies to a single detector. The generalization to a detector network is provided in Appendix~A of Ref.~\cite{Kapil:2024zdn}.}
\begin{equation}
\label{bias}
    \Delta \theta^{\rm rec}_i \equiv (\Gamma^{\rm rec}_{ij})^{-1} \left\langle \frac{\partial h_{\rm rec}}{\partial \theta_j^{\rm rec} } \bigg\vert h_{\rm true} - h_{\rm rec} \right\rangle\,.
\end{equation}
We compute the systematic errors~\eqref{bias} for three scenarios with different combinations of true and recovery signals, summarized in Table~\ref{tab:res}. This analysis will be used to identify the values of $\epsilon$ and $\tilde{L}$ that induce the largest bias in the source parameters. We compute the Fisher matrices and Cutler-Vallisneri biases through the \texttt{gwfast} package~\cite{Iacovelli:2022mbg}. Notice that the bias is evaluated only after performing the waveform alignment procedure described, e.g., in Refs.~\cite{Dhani:2024jja,Kapil:2024zdn}. In particular, we compute the values of the coalescence time and phase for the recovery waveform by maximizing the overlap 
\begin{equation}\label{eq:overlap}
    {\cal O} (h_{\rm true}, h_{\rm rec}) = \dfrac{\langle h_{\rm true} | h_{\rm rec}\rangle}{\sqrt{\langle h_{\rm true} | h_{\rm true}\rangle \langle h_{\rm rec} | h_{\rm rec}\rangle}}\,,
\end{equation}
and use those to evaluate Eq.~\eqref{bias}. For improved numerical performance and accuracy, the coalescence phase resulting in the maximum overlap is found analytically (see e.g. Sec.~IV of Ref.~\cite{Allen:2005fk}), while for the coalescence time we resort to a numerical optimization through \texttt{scipy}~\cite{2020SciPy-NMeth}. We found that the alignment procedure only marginally affects our results. This is expected, as the adopted waveform models are modifications of the same approximant, so they adopt the same conventions.

\begin{figure*}[tbp] 
    \includegraphics[width=0.8\textwidth]{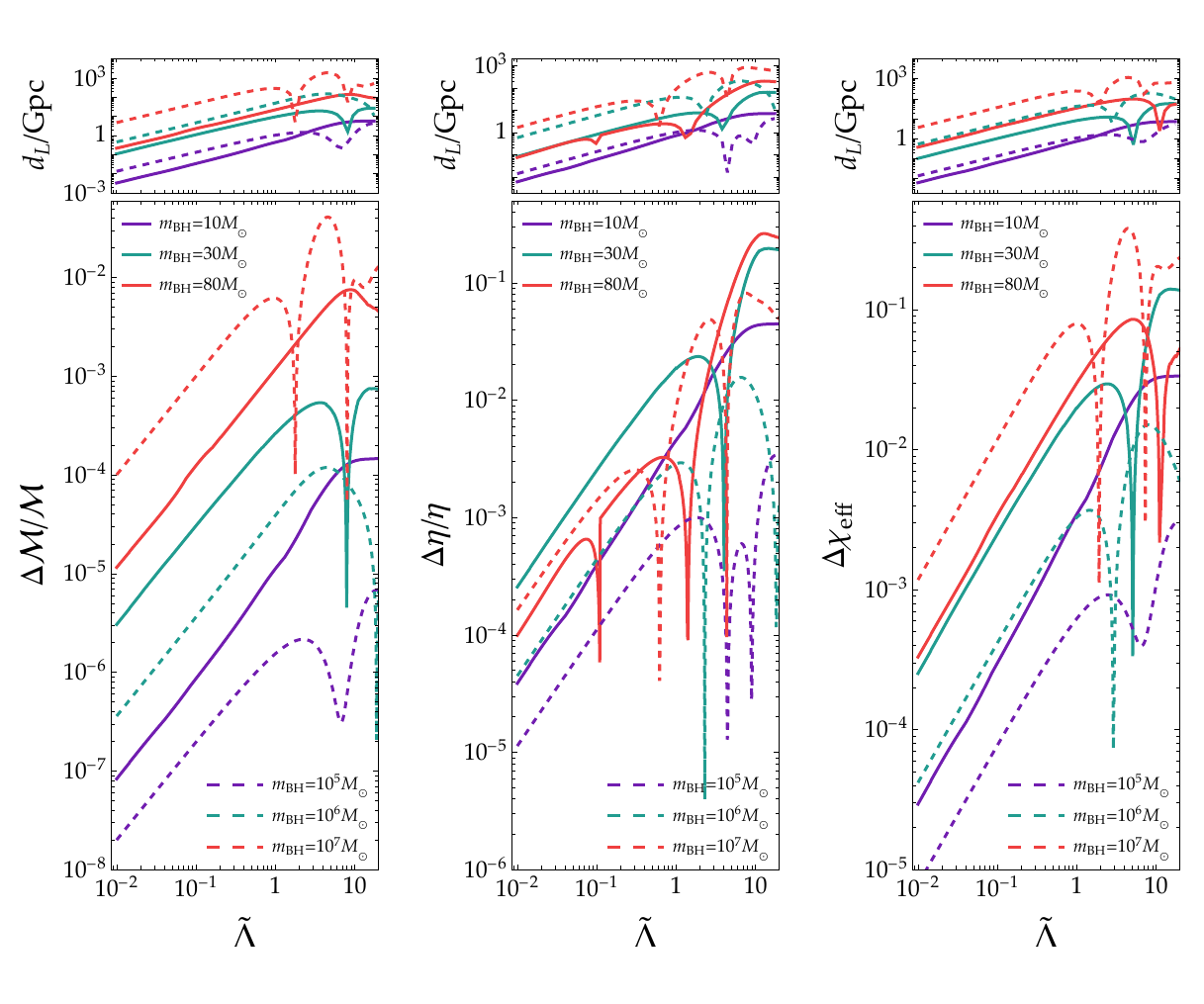}
    \caption{Lower panels: $\Delta \mathcal{M}/\mathcal{M}$ (left), $\Delta \eta/\eta$ (center), and $\Delta \chi_{\rm eff}$ (right), computed for case~1, as a function of the tidal deformability. Here  $\Delta \theta_i$ denotes the systematic error on the corresponding parameter. Solid and dashed curves in each panel refer to equal-mass binaries with $m_\text{\tiny BH} = (10,\, 30,\, 80)\, {\rm M_\odot}$ and $m_\text{\tiny BH} = (10^5,\, 10^6,\, 10^7)\, {\rm M_\odot}$ observed by ET and LISA, respectively. Upper panels: luminosity distance at which the systematic error becomes equal to the statistical uncertainty for each binary.}
\label{fig:case1}
\end{figure*}

\section{Numerical results}\label{Sec:num}

We discuss the effect of systematics by considering, for each of the three cases listed in Table~\ref{tab:res}, two sets of mass configurations: $m_\text{\tiny BH}=(10,\,30,\,80)\,{\rm M_\odot}$ for ET, and $m_\text{\tiny BH}=(10^5,\,10^6,\,10^7)\,{\rm M_\odot}$ for LISA. These correspond to SNRs of $\rho\sim\!(210,\, 500,\, 1050)$ for ET and SNRs of $\rho\sim\!(1720,\, 10070,\, 5890)$ for LISA, respectively, at a distance of $d_L=1\,{\rm Gpc}$.

For cases 1 and 3, where tidal effects persist throughout the inspiral phase for some of the considered waveform models, we restrict our analysis to $\tilde{\Lambda} \lesssim 20$ to ensure that $\psi_{\rm tidal} \lesssim \psi_\text{\tiny PP}$ at the ISCO frequency. This guarantees that the leading $5\,{\rm PN}$ tidal phase contribution remains subdominant to the point-particle term across all frequencies. Our constraint on $\tilde{\Lambda}$ is conservative; the waveform models we adopt may still provide a faithful description of the signal for larger $\tilde{\Lambda}$ at frequencies below merger~\cite{Chia:2023tle}. In Appendix~\ref{app-B} we will show the results of Case~1 when relaxing this assumption.
For cases in which environmental effects fade progressively during the inspiral, we consider larger values of the tidal deformability, since $\tilde{\Lambda}$ vanishes at frequencies small enough to avoid any potential breaking of the PN expansion.

\subsection{Case~1: vacuum vs vacuum\texorpdfstring{\,}{ }+\texorpdfstring{\,}{ }TLN}

We begin with the simplest setup, denoted as case~1 in Table~\ref{tab:res}, where the true waveform is characterized by a constant, nonzero tidal deformability throughout the entire inspiral. This scenario describes physical systems in which tidal effects persist during the coalescence, e.g. binary mergers involving neutron stars~\cite{Shibata:2001ag,Marronetti:2003hx,Dietrich:2015pxa}, exotic compact objects, or BHs in modified gravity theories.
As the recovery signal, we use a pure vacuum template. Schematically:
\begin{equation}
\label{case1}
    \begin{aligned}
        h_{\rm rec} & = h(f,\bm{\theta}\setminus \{\tilde{\Lambda}\})|_{\mathcal{S},\mathcal{S}_m = 1}\ , \\ 
        h_{\rm true} & = h(f,\bm{\theta})|_{\mathcal{S},\mathcal{S}_m = 1}\ ,
    \end{aligned}
\end{equation}
where $h(f)$ is the full environmental waveform discussed in the previous section but neglecting the fading effect, i.e., $\mathcal{S} = 1$ and $\mathcal{S}_m = 1$, and the symbol $\bm{\theta}\setminus \{\tilde{\Lambda}\}$ indicates that the TLN $\tilde{\Lambda}$ is not included in the model's parameter array $\bm{\theta}$. We inject these waveform models into Eq.~\eqref{bias}, computing the bias on the source parameters.

The lower panels of Fig.~\ref{fig:case1} show
$\Delta \mathcal{M}/\mathcal{M},\, \Delta \eta/\eta$, and $\Delta \chi_{\rm eff}$ as functions of $\tilde{\Lambda}$.
For both detectors, the bias generally increases with $\tilde{\Lambda}$.  This trend is expected, as the phase difference between the recovery signal and the true signal is proportional to $\tilde{\Lambda}$, and grows secularly if the tidal deformability is constant. The behavior of the bias shown in Fig.~\ref{fig:case1} can also be understood more quantitatively through a simple scaling estimate~\cite{Favata:2021vhw}, assuming negligible correlations among the waveform parameters: 
\begin{equation}
    \begin{aligned}
        \Delta \theta_i &\simeq  (\Gamma^{\rm rec}_{ij})^{-1} \left\langle \frac{\partial h_{\rm rec}}{\partial \theta_j^{\rm rec} } \bigg\vert  h_{\rm rec}\, {\rm i}\, \psi_\text{\rm tidal} \right\rangle \\ 
        &\simeq \rho \, \sigma_i \psi_\text{\rm tidal} (f_\text{\tiny min})\,,
    \end{aligned}
\label{eq:cas0scaling}
\end{equation}
where $\psi_{\rm tidal}\propto \tilde{\Lambda}$ is evaluated at the minimum frequency $f_\text{\tiny min}$, at which the integral in the scalar product is peaked. Equation~\eqref{eq:cas0scaling} supports the numerical finding that $\Delta\theta_i$ increases for larger TLNs.

The bias on the chirp mass is always much smaller than its injected value, though $\Delta{\cal M}$ can become $\sim\!10\%$ of ${\cal M}$ for $\tilde{\Lambda}\gg1$ and for the heaviest binaries we consider in the LISA band. Indeed, given that the signal's amplitude and statistical uncertainties increase with the binary's total mass, Eq.~\eqref{eq:cas0scaling} shows that $\Delta \theta_i$ is expected to grow for heavier binaries, as evident from all panels in Fig.~\ref{fig:case1}. 

Note that $\Delta\eta/\eta$ is generally larger than $\Delta \mathcal{M}/\mathcal{M}$, with the bias exceeding $10\%$ of the injected symmetric mass ratio for BHs with $m_\text{\tiny BH}=80\,{\rm M}_{\odot}$. Moreover, the systematic error introduced by tidal effect mismodeling is particularly significant for the effective spin parameter, given the injected value $\chi_{\rm eff}=0$. As shown in the right panel of Fig.~\ref{fig:case1}, this bias can reach up to $\Delta\chi_{\rm eff}\simeq \mathcal{O}(0.1)$ for both ET and LISA binaries.

The oscillations in $\Delta\theta_i$ observed at large values of  $\tilde{\Lambda}$, and common to all panels of Fig.~\ref{fig:case1}, are due to the magnitude of the tidal phase in Eq.~\eqref{bias}. For large values of the tidal deformability, the difference  $h_{\rm true}-h_{\rm rec}\sim (e^{{\rm i}\psi_{\rm tidal}} - 1)$ cannot be linearized around $\tilde{\Lambda}$, leading to an oscillatory behavior.
Let us emphasize that, in such cases, the template-signal mismatch becomes sufficiently large that a real detection would likely reveal substantial fitting issues in the data analysis, rather than merely minor systematic errors in parameter estimation, ultimately preventing any meaningful recovery of the source parameters. This highlights that the results obtained for large values of  $\tilde{\Lambda}$ should be viewed with caution when interpreted as systematic biases. A similar comment will also apply when considering large environmental scales $\tilde{L}$ in Case~3.

To further assess the impact of bias on the overall error budget of source parameters, we determine the luminosity distance at which the systematic error, $\Delta\theta_i$, equals the statistical uncertainty, $\sigma_{i}$, as estimated by the Fisher matrix. From the scaling in Eq.~\eqref{eq:cas0scaling}, we have $\Delta\theta_i/\sigma_{i} \sim \tilde{\Lambda}/d_L$, suggesting that the distance at which this ratio reaches unity scales linearly with the tidal deformability. This trend is confirmed in the top panels of Fig.~\ref{fig:case1}, which show the values of $d_L$ for which $\Delta\theta_i/\sigma_i=1$ for the same binaries analyzed in the lower panels.

Using the tidal phase expression from Eq.~\eqref{eq:5PN} together with Eq.~\eqref{eq:cas0scaling}, we find that imposing this condition on the chirp mass implies
\begin{equation}\label{eq:case0_scaling}
    \begin{aligned}
        &\frac{\Delta \mathcal{M}}{\sigma_{\mathcal{M}}} \gtrsim 1 \,\,\, \implies \\
        &\dfrac{d_L}{{\rm Gpc}}  \lesssim\!\left\{
        \begin{aligned}
            &10\!\left( \dfrac{f_\text{\tiny min}}{2 \, \rm Hz} \right)^{\!\!\nicefrac{1}{2}}\!\! \left( \dfrac{m_\text{\tiny BH}}{30\, {\rm M_\odot}} \right)^{\!\!\nicefrac{5}{2}}\!\!\tilde{\Lambda} \ \text{for ET}\,, \\
            &50\!\left( \dfrac{f_\text{\tiny min}}{10^{-5} \, \rm Hz} \right)^{\!\!\nicefrac{1}{2}} \!\!\left( \dfrac{m_\text{\tiny BH}}{10^6\, {\rm M_\odot}} \right)^{\!\!\nicefrac{5}{2}}\!\!\tilde{\Lambda} \ \text{for LISA} \,.
        \end{aligned}\right.
    \end{aligned}
\end{equation}
Equation~\eqref{eq:case0_scaling} also indicates that, at a fixed luminosity distance, larger masses require smaller TLNs for the bias to dominate over the statistical error. This scaling holds qualitatively for $\eta$ and $\chi_{\rm eff}$ as well.

The results for case~1 have important implications for measuring tidal effects in different astrophysical scenarios, such as mergers of BH mimickers (exotic stellar configurations with a rigid surface instead of an event horizon~\cite{Carballo-Rubio:2025fnc}). The TLN of such objects can exhibit a logarithmic dependence on compactness~\cite{Cardoso:2017cfl}. In this case, $\tilde{\Lambda}$ can reach values as high as $10^{-2}$, even when deviations from the BH geometry occur only a Planck length away from the would-be horizon~\cite{Pani:2015tga, Cardoso:2017cfl}.  

Equation~\eqref{eq:case0_scaling} shows that mimickers with ${m_\text{\tiny BH}} \sim 30 \, { {\rm M}_\odot}$ (${m_\text{\tiny BH}} \sim 5\times 10^5 \, { {\rm M}_\odot}$) observed by ET (LISA) with $\tilde{\Lambda} \sim 10^{-2}$ at $d_L \lesssim 10 \, \rm{Gpc}$ would introduce a significant bias in parameter inference if interpreted as BHs (i.e., assuming $\tilde{\Lambda} = 0$).  

\begin{figure}[tbp] 
    \includegraphics[width=\columnwidth]{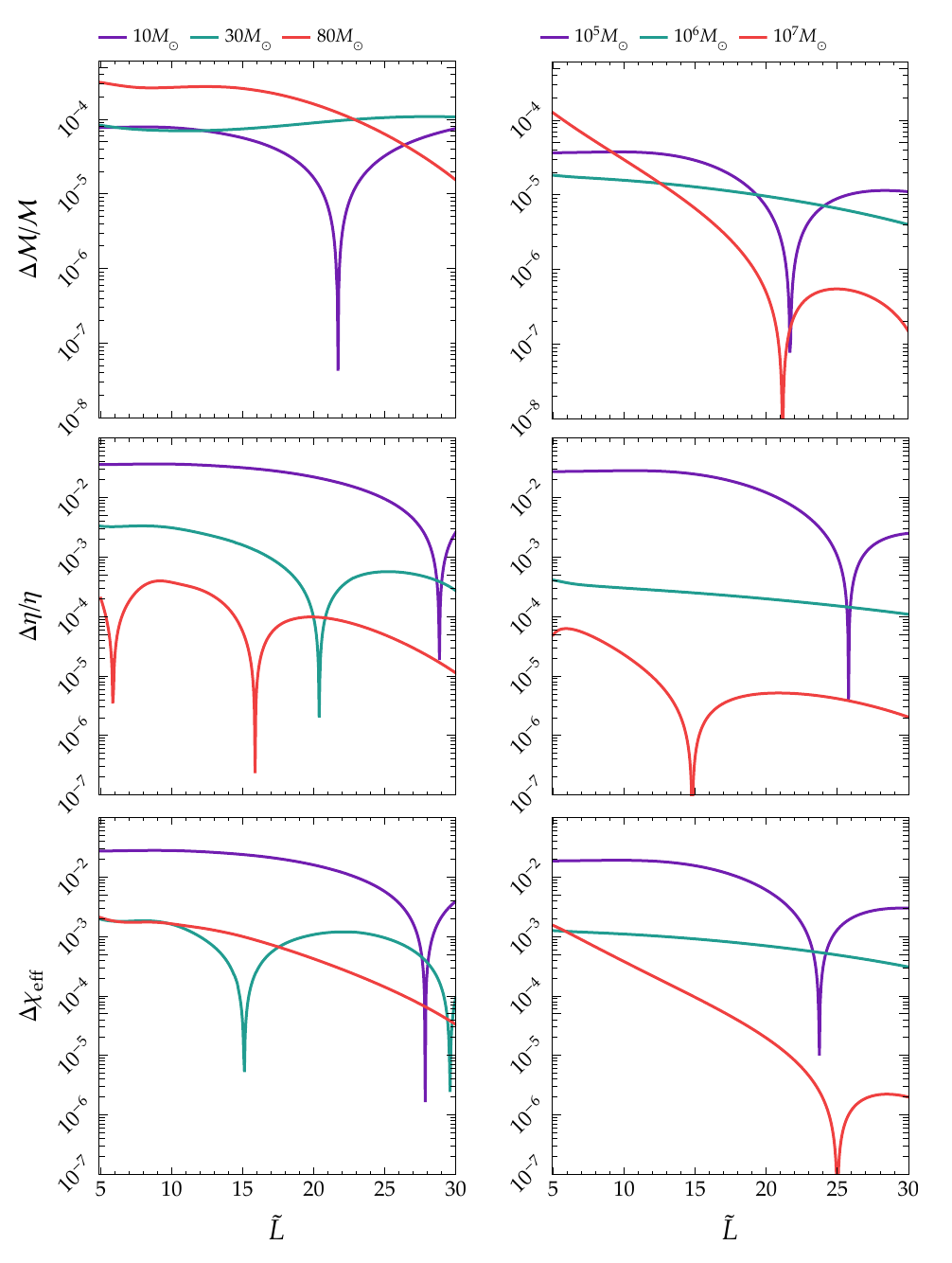}
    \caption{Ratio between the systematic error and the injected value for the chirp mass (top row), the symmetric mass ratio (middle row), and the effective spin parameter (bottom row), computed for case~2, as a function of $\tilde{L}$. We consider BHs with masses $m_\text{\tiny BH} = (10,\, 30,\, 80)\, {{\rm M}_\odot}$ and $m_\text{\tiny BH} = (10^5,\, 10^6,\, 10^7)\, { {\rm M}_\odot}$ as observed by ET (left panels) and LISA (right panels), respectively. We also assume $\epsilon=10^{-3}$.}
\label{fig:case2}
\end{figure}

\subsection{Case~2: vacuum vs fading environment}

In case~2, the true GW signals correspond to BHs surrounded by a fading environment, where both the BH masses and TLNs evolve as the inspiral progresses and the frequency increases. Once the cut-off frequency $f_\text{\tiny cut} \propto \tilde{L}^{\nicefrac{-3}{2}}/m_\BH$ is reached, the binary transitions to a vacuum configuration, in which $\tilde{\Lambda}$ vanishes and the BH masses revert to their bare values. We recover these signals using vacuum templates, namely: 
\begin{equation}
    \begin{aligned}
    \label{case2}
        h_{\rm rec} & = h(f,\bm{\theta} \setminus \{\tilde{\Lambda}\})|_{\mathcal{S},\mathcal{S}_m = 1}\,, \\ 
        h_{\rm true} & = h(f, \bm{\theta})\,.
    \end{aligned}
\end{equation}
Hereafter, we fix the $\epsilon$ factor, which appears in the frequency-dependent mass~\eqref{smoothingM} and tidal deformability~\eqref{eq:tidalparamet}, to $\epsilon = 10^{-3}$.
This choice of $\epsilon$ is representative of realistic astrophysical environments with matter overdensities around BHs~\cite{Abramowicz:2011xu, Inayoshi:2019fun}. For instance, it corresponds to the maximum matter content in thin accretion disks beyond which precession effects become significant~\cite{Tiede:2023cje}, or to a conservative value for the mass of a scalar cloud formed through superradiance~\cite{Brito:2015oca}.

Figure~\ref{fig:case2} shows the bias in ${\cal M},\, \eta$, and $\chi_{\rm eff}$ as a function of the effective environmental scale $\tilde{L}$ for ET and LISA binaries.
An extension of this analysis in which the TLN is included as a parameter of the model is left to Appendix~\ref{app-C}.
A large $\tilde{L}$ impacts the waveform through two competing effects: (\emph{i}) it increases $\tilde{\Lambda}$, enhancing the amplitude of the tidal phase, and (\emph{ii}) it lowers the cut-off frequency, causing the environmental effects to fade away earlier in the inspiral.

Similar to case~1, our analysis shows that the ratio $\Delta \theta_i/\theta_i$ remains much smaller than one across the entire parameter space we consider for the chirp mass. However, it becomes significant for both the symmetric mass ratio and the effective spin. The bias is strongly dependent on $\epsilon$. We computed $\Delta \theta_i$ for different values of this parameter (not shown here) and found that systematic errors scale linearly with $\epsilon$. 
A note of caution is needed: while biases on ${\cal M}$ can be small compared to ${\cal M}$ itself, this may not be the case when we compare them with the statistical uncertainty. Because $\cal M$ is the best-constrained mass parameter in the inspiral, uncertainties for ET and LISA can be as low as $\sigma_{\cal M}/{\cal M}= {\cal O}(10^{-5})$ and $\sigma_{\cal M}/{\cal M} = {\cal O}(10^{-5}-10^{-6})$, respectively, for sources at $d_L=1\,{\rm Gpc}$. Biases can thus be 10 times larger than the statistical uncertainty, especially for low $\tilde{L}$. We emphasize that, for values of $\tilde{L} \lesssim \mathcal{O}(5)$, the environment would neither be in equilibrium nor gravitationally bound to the BHs, as discussed in Sec.~\ref{sec:env}. In other words, in this regime, the component BHs would not be embedded in an external environment. While our analysis is restricted to $\tilde{L} \gtrsim \mathcal{O}(5)$ -- where the presence of an external environment leads to an augmented mass and nonvanishing TLNs, and hence a nonzero bias -- we do not model the environmental physics in detail at smaller scales $\tilde{L}$. In this limit, the inferred bias is generally expected to vanish, consistently with the absence of environmental effects (even though the very definition of bias may become ambiguous, as the system becomes fully degenerate with a BH binary of component masses $m_\BH (1 +\epsilon)$). This conclusion can be affected by the presence of strong correlations among the environmental and vacuum model parameters, which could possibly cause  inference bias even when these effects are absent in the actual signal~\cite{Kejriwal:2023djc}.

The systematic errors in Fig.~\ref{fig:case2} exhibit a different trend with the scale $\tilde{L}$ (and thus with the TLN) compared to case~1. This behavior can be understood by analyzing the dependence of our calculations on the cut-off frequency, $f_\text{\tiny cut} (m_\BH, \tilde{L})$. First, note that the true waveform accumulates a mismatch with the recovery signal only for $f < f_\text{\tiny cut}$. This implies that, for a fixed BH mass, larger values of $\tilde{L}$ introduce a bias as long as the cut-off frequency remains above the minimum frequency of integration for the given GW detector, i.e., $f_\text{\tiny cut} \gtrsim f_\text{\tiny min}$. This effect is particularly evident for the lightest BH shown in Fig.~\ref{fig:case2}, where $f_\text{\tiny cut}$ reaches higher values, extending the frequency range over which the mismatch accumulates.
For more massive BHs, $f_\text{\tiny cut}$ shifts to lower frequencies, causing the waveform to transition to the vacuum configuration earlier, thereby reducing the bias. Specifically, for $m_\BH = 80 \, {\rm M}_\odot$ in the ET case and $m_\BH = 10^7 \, {\rm M}_\odot$ in the LISA case, the cut-off frequency approaches the detector’s minimum frequency when $\tilde{L} \approx 30$, resulting in a rapid decrease in bias at that point.

Unlike case~1, the bias shown in Fig.~\ref{fig:case2} arises from changes in both $\tilde{\Lambda}$ and the BH masses, with the latter contributing the most. Even a small variation in $m_\BH$ of the order of $10^{-3}$ accumulates over the many cycles the binary spends in the detector's low-frequency band (we expect this effect to become even more relevant for lighter binaries, as those in the subsolar mass range potentially observable by ET~\cite{Barsanti:2021ydd, Franciolini:2021xbq, Franciolini:2023opt, Crescimbeni:2024cwh, DeLuca:2024uju}). This effect tends to dominate over the contribution from tidal deformability, which becomes more significant at higher frequencies, where the number of GW cycles is smaller.

Using Eq.~\eqref{bias} and noting that the waveform difference $h_{\rm true} - h_{\rm rec}$ is now primarily driven by the mass correction to the leading $0\,{\rm PN}$ term in the point-particle phase, we can find a semi-analytic expression for the bias:
\begin{equation}
    \Delta \theta_i \sim \rho \, \sigma_{i} \delta \psi_\text{\rm 0PN} (f_\text{\tiny min})\label{eq:cas1scaling}\ ,
\end{equation}
where, to leading order in $\epsilon$, the phase change due to mass evolution is given by
\begin{equation}
    \delta \psi_\text{\rm 0PN} (f) \simeq \frac{5}{128 \eta} (\pi M f)^{\nicefrac{-5}{3}} \epsilon \,.\label{eq:cas1scaling2}
\end{equation}
Equations~\eqref{eq:cas1scaling} and \eqref{eq:cas1scaling2} allow us to establish a condition on the values of $\epsilon$ for which systematic and statistical uncertainties become comparable. Neglecting correlations among the waveform parameters we find: 
\begin{equation}
  \frac{\Delta \theta_i}{\sigma_{i}} \gtrsim 1 \,\implies \,\frac{\epsilon}{10^{-3}} \gtrsim \frac{\mathcal{O}(10^3)}{\rho}\left(\frac{f_\text{\tiny min}\, m_\text{\tiny BH}}{10^{-4}} \right)^{\nicefrac{5}{3}} \,.\label{eq:cas1scaling2eps}
\end{equation}
The constraint~\eqref{eq:cas1scaling2eps} highlights two key results of the case~2 analysis. First, at leading order in the matter overdensity $\epsilon$, the bias is independent of the environmental scale $\tilde{L}$ (mostly in the small $\tilde{L}$ regime), since the tidal phase contribution is subdominant in the difference between the true and recovered waveforms (notice that large values of $\tilde{L}$ affect the number of cycles over which a mismatch can be accumulated, and thus the corresponding bias). Second, heavier BHs require larger values of $\epsilon$ for the bias to exceed the statistical error. Similarly to Eq.~\eqref{eq:case0_scaling}, we can derive the luminosity distance at which the two errors become comparable to be 
\begin{eqnarray}
\label{eq:case1_scaling}
\dfrac{d_L}{{\rm Gpc}}  \lesssim\!\left\{
        \begin{aligned}
            &\!\left( \dfrac{f_\text{\tiny min}}{2 \, \rm Hz} \right)^{\!\!\!\!\nicefrac{-17}{6}}\!\! \left( \dfrac{m_\text{\tiny BH}}{30\, {\rm M_\odot}} \right)^{\!\!\!\!\nicefrac{-5}{6}}\!\!\left( \dfrac{\epsilon}{10^{-3}} \right) \ \text{for ET}\,, \\
            &4\!\left( \!\dfrac{f_\text{\tiny min}}{10^{-5} \, \rm Hz} \!\right)^{\!\!\!\!\nicefrac{-17}{6}} \!\!\!\left(\! \dfrac{m_\text{\tiny BH}}{10^6\, {\rm M_\odot}} \!\right)^{\!\!\!\!\nicefrac{-5}{6}}\!\!\!\left( \dfrac{\epsilon}{10^{-3}} \right)\ \text{for LISA} \,.
        \end{aligned}\right.
\end{eqnarray}
This expression accurately reproduces our numerical results for masses smaller than $30\, {\rm M_\odot}$ for ET ($10^6\, {\rm M_\odot}$ for LISA) in the small $\tilde{L}$ regime, whereas it becomes conservative for larger masses. For comparison, the numerical computed value of $d_L$ for $10\, {\rm M_\odot}$  ($10^5\, {\rm M_\odot}$) is $d_L \approx 3 \, {\rm Gpc}$ ($d_L \approx 24 \, {\rm Gpc}$).

\subsection{Case~3: fixed environment vs fading environment}

In case~3, the waveform mismatch is given by a recovery waveform that does not take into account changes in the environment during the inspiral, i.e., the masses and tidal parameters of the recovery waveform are fixed to their asymptotic values far from the merger.
The templates for case~3 are given by 
\begin{equation}
    \begin{aligned}
    \label{case3}
        h_{\rm rec} & = h(f,\bm{\theta})|_{\mathcal{S} = 1 ,\mathcal{S}_m = 1 + \epsilon, \tilde{\Lambda} = \frac{2}{3}\epsilon \,\tilde{L}^5}\,, \\ 
        h_{\rm true} & = h(f,\bm{\theta})\,,
    \end{aligned}
\end{equation}
where the smoothing functions are set to their value at the beginning of the inspiral phase, $ {\cal S}_{m} = {\cal S}_{m} (f_\text{\tiny min}) \approx 1 + \epsilon$ and  ${\cal S} = {\cal S}(f_\text{\tiny min}) \approx 1$, such that $h_{\rm rec}$ is augmented by a fixed dress.

\begin{figure}[tbp] 
    \includegraphics[width=\columnwidth]{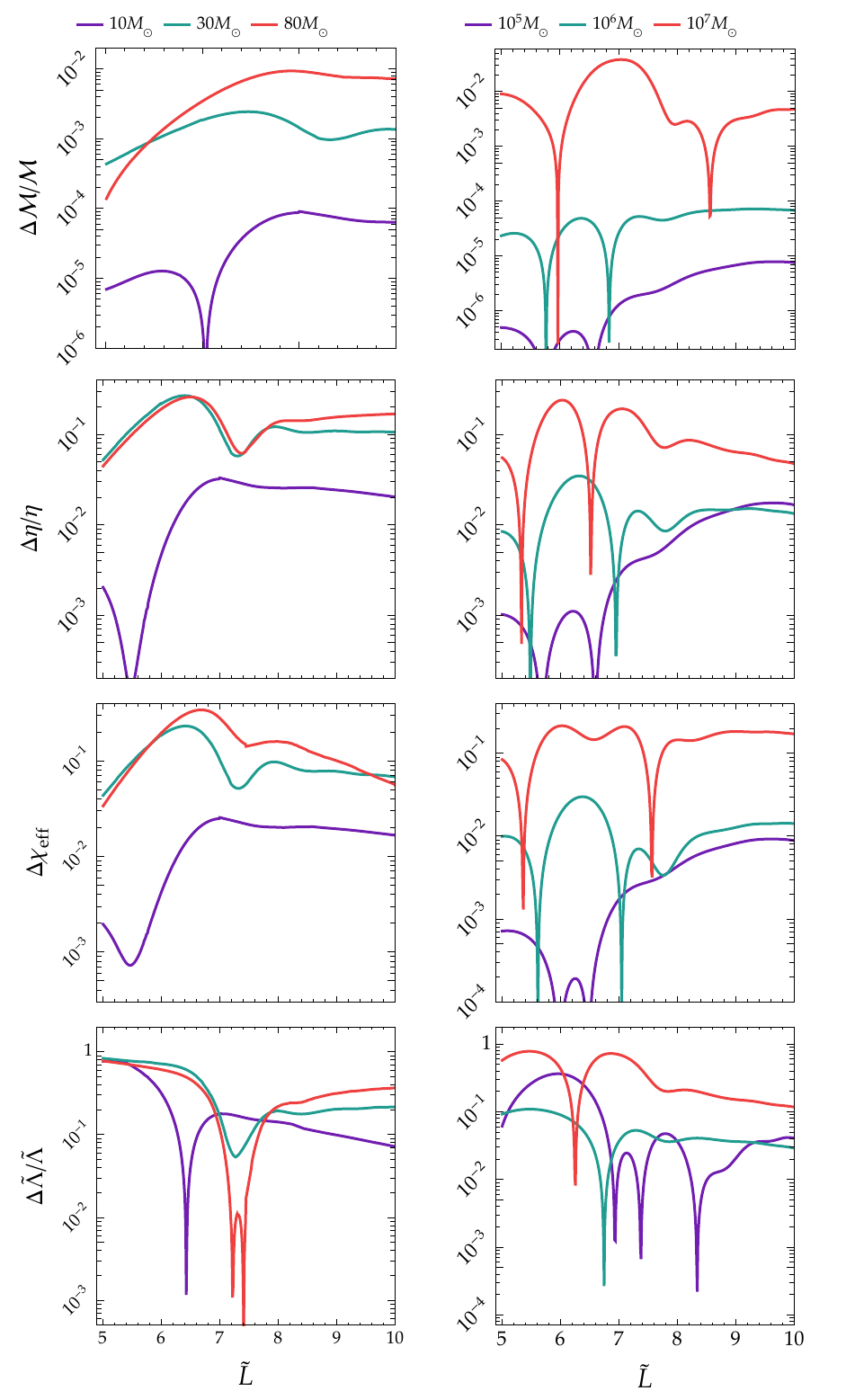}
    \caption{Same as Fig.~\ref{fig:case2} but for case~3, schematically described in Eq.~\eqref{case3}. All results shown in the panels correspond to binary configurations with $\epsilon=10^{-3}$.}
    \label{fig:case3}
\end{figure}

The results for the bias are shown in Fig.~\ref{fig:case3}. Along with ${\cal M},\,\eta$ and $\chi_{\rm eff}$ we also show systematic errors on the tidal deformability, which is now included within the source parameters of the recovery template. 

The magnitude of the ratio between $\Delta\theta_i$ and the injected values follows the same hierarchy observed in cases~1 and~2, with the chirp mass being the parameter least affected by waveform mismatch. Systematic errors on the symmetric mass ratio and the effective spin parameter cluster between approximately $1\%-10\%$ for the heaviest binaries we consider, both in ET and LISA, and they tend to be smaller for lighter BHs. However, the bias on $\tilde{\Lambda}$ is significant:  $\Delta\tilde{\Lambda}/\tilde{\Lambda}$ reaches values of the order of $10\%$ and higher, independent of the BH mass and the effective scale $\tilde{L}$. 

Unlike previous cases, this bias is purely a \emph{high-frequency} effect, induced during the late inspiral phase after the binary reaches the Roche frequency. The recovery waveform fails to account for changes in $m_\text{\tiny BH}$ and $\tilde{\Lambda}$, leading to systematic discrepancies. Indeed, the tidal term provides the leading contribution to the waveform difference,  $h_{\rm true}-h_{\rm rec}\sim (e^{{\rm i}\psi_{\rm tidal}} - 1)$, giving rise to the highly oscillatory behavior observed in the plot and to higher biases for heavier BH masses. Note that for $\tilde{L} \in (6 - 10)$ the TLN $\tilde{\Lambda}_{\rm rec} \simeq \epsilon \, \tilde{L}^5$ is already sizable, so that the phase $e^{{\rm i}\psi_{\rm tidal}}$ cannot be linearized in $\tilde{\Lambda}$, thus confirming the  oscillatory trends.

In contrast to case~2, we expect the systematic error to become even more significant for values of $\tilde{L}$ larger than those considered in Fig.~\ref{fig:case3}. Increasing the environmental scale will cause the masses and TLN to fade earlier in $h_{\rm true}$, thereby increasing the mismatch at a constant rate, similar to case~1. However, as discussed before, we restrict the range to $\tilde{L}\lesssim 10$ to ensure the validity of the point particle series expansion.

For the reference binaries located at $d_L=1\,{\rm Gpc}$ employed earlier, the systematic uncertainties can be ${\cal O}(10-100)$ times larger than the corresponding statistical uncertainties on the chirp mass at both ET and LISA, with the larger masses showing higher biases. For the tidal deformability, systematics ${\cal O}(10-10^3)$ larger than statistical uncertainties can be found, again especially for the more massive systems in LISA.

\section{Conclusions}
\label{sec:conc}

Binary BH systems are typically expected to evolve in vacuum. However, secular effects such as baryonic accretion, dark matter clouds, or superradiance, may induce the presence of an external environment surrounding the binary components. This environment is expected to influence the properties and dynamics of the binary, significantly affecting how the binary is interpreted in GW data analysis and potentially altering conclusions about the physical characteristics of the system.

In this study, we assess the systematic biases in GW observations for upcoming experiments (like ET and LISA) which occur when events are assumed to take place in a vacuum, disregarding the influence of environmental tidal interactions. To do so, we have focused on compact enough environmental clouds surrounding each component in the binary, thus leading to corrections to their masses and sourcing sizable TLNs. These effects are anticipated to alter the binary evolution until the Roche radius is reached, at which point tidal interactions become significant enough to disrupt the component clouds, leading to naked binary components in the last stages of the coalescence.

Throughout the paper, we have made several comparisons between true environmental and recovery waveforms, based on the mismatch induced by neglecting environmental effects, which are briefly summarized in Table~\ref{tab:res}. The main results of such comparisons are the following. While the systematic uncertainties for the binary chirp mass are usually smaller than the injected values, such errors can be dramatically more important for the other model parameters, such as the symmetric mass ratio, the effective spin parameter, and the tidal deformabilities. This conclusion is particularly relevant for environments surrounding the binary components with high density (high $\epsilon$).
The systematic uncertainty can be comparable to or larger than the statistical one for large SNR events.
Let us also highlight that our results are conservative: the use of more refined waveform models, such as \textsc{IMRPhenomHM}~\cite{London:2017bcn}, which includes the contribution of higher-order harmonics, would lead to binary parameter measurements with higher statistical accuracy (see e.g. Fig.~6 of Ref.~\cite{Iacovelli:2022bbs}), resulting in larger mismatches between the true and recovery waveforms and leading to larger biases.

There are numerous avenues for future research. A key objective is to expand our analysis by incorporating more sophisticated models of environmental effects, ranging from accretion disks to superradiance clouds, in order to gain a deeper understanding of the induced systematic biases. In relation to this point, note that an external environment may also induce (small) corrections to the tidal heating dissipative contribution in the BH response, which enter the waveform at a lower PN order~\cite{Chia:2024bwc}, and thus have to be taken into account for a more complete analysis.
Additionally, generalizing our results through full Bayesian analyses would provide valuable insight into the reconstruction of model parameters in the presence of dense environments surrounding binary BH systems, along the line of Refs.~\cite{CanevaSantoro:2023aol,Roy:2024rhe}. Finally, we aim to conduct an in-depth comparison of environmental effects with modified gravity theories within the context of tidal searches. These extensions will be explored in future work.

\begin{acknowledgments}
We thank Daniel D'Orazio, Konstantinos Kritos, Toni Riotto and Jay Wadekar for interesting discussions. 
We also thank the anonymous referee for their valuable feedback on the manuscript.
V.D.L. thanks the Johns Hopkins University for the warm hospitality during the realization of this project. 
V.D.L. is supported by funds provided by the Center for Particle Cosmology at the University of Pennsylvania.
E.B., L.D.G. and F.I. are supported by NSF Grants No.~PHY-2207502, AST-2307146, PHY-090003 and PHY-20043, by NASA Grant No.~21-ATP21-0010, by the John Templeton Foundation Grant 62840, and by the Simons Foundation.
L.D.G. is supported by Simons Investigator Grant No. 144924.
The work of F.I. is supported by a Miller Postdoctoral Fellowship. 
E.B. and A.M. acknowledge support from the ITA-USA Science and Technology Cooperation program, supported by the Ministry of Foreign Affairs of Italy (MAECI).
A.M. acknowledges financial support from MUR PRIN Grants No.~2022-Z9X4XS and No.~2020KB33TP. 
\end{acknowledgments}

\appendix

\section{Interactions with surrounding matter after 
the fading}
\label{app-A}

The tidal disruption of the BH dress during coalescence is expected to pollute the binary surroundings. 
One might argue that the environmental debris could influence the system's evolution by interacting with the binary components, in a manner analogous to circumbinary structures (see the right panel of Fig.~\ref{fig:setup} for a schematic illustration). While a comprehensive understanding of such dynamics likely requires detailed simulations, here we aim to provide analytical insights demonstrating that the dynamics after tidal disruption is expected to be dominated by GW emission. Specifically, we consider several key mechanisms that could impact the evolution of the BHs -- namely, matter accretion, dynamical friction, and drag forces -- and show that the associated energy losses are subdominant compared to those from gravitational radiation.

First of all, we roughly estimate the BH accretion rate onto the binary as~\cite{Shapiro:1983du} 
\begin{equation}\label{eq:accretion}
    \dot{m}_{\text{\tiny{BH}}} \sim \rho_\text{\tiny env} \, m_{\text{\tiny{BH}}}^2 /v^3 \sim \epsilon/\tilde{L}^{\nicefrac{3}{2}} \,,
\end{equation}
where $\rho_\text{\tiny env} \sim  m_{\text{\tiny env}}/d_{\text{\tiny{Roche}}}^3$ is the ambient density, and $v \sim (\pi \,m_{\text{\tiny{BH}}}f_{\text{\tiny{cut}}} )^{\nicefrac{1}{3}}$ is the binary velocity. The characteristic accretion timescale is proportional to the Salpeter time and is given by $t_{\text{\tiny{acc}}} \sim m_{\text{\tiny{BH}}} \tilde{L}^{\nicefrac{3}{2}}/\epsilon$. On the other hand, the orbital period after the tidal disruption event could be estimated as $T_\text{\tiny GW} \sim  1/f_{\text{\tiny{cut}}}  \sim m_{\text{\tiny{BH}}} \tilde{L}^{\nicefrac{3}{2}}$. Their ratio thus reads
\begin{equation}
    \frac{t_{\text{\tiny{acc}}}}{T_\text{\tiny GW}} \sim   \frac{1}{\epsilon} \,.
\end{equation}
For realistic models with $\epsilon \lesssim 10^{-2}$, we can easily conclude that accretion is negligible.

The second effect we consider is dynamical friction. In this case, the energy loss by the binary BH system after the fading of the environment (assuming almost collisionless particles in the medium) is roughly given by~\cite{Chandrasekhar:1943ys} 
\begin{equation}
    \dot{E}_{\text{\tiny{dyn}}}\sim \rho_\text{\tiny env} \, m_{\text{\tiny{BH}}}^2 /v \sim \epsilon/\tilde{L}^{\nicefrac{5}{2}} \, .
\end{equation}
On the other hand, the energy loss through the emission of GWs is estimated to be $\dot{E}_{\text{\tiny{GW}}}\sim m_\text{\tiny BH}^2 v^6/d_\text{\tiny Roche}^2 \sim 1/\tilde{L}^{5}$~\cite{Peters:1963ux, Peters:1964zz}. 
Therefore, their ratio reads
\begin{equation}
    \frac{\dot{E}_{\text{\tiny{dyn}}}}{\dot{E}_{\text{\tiny{GW}}}} \sim \epsilon\, \tilde{L}^{\nicefrac{5}{2}}\,.
\end{equation}
For the benchmark value $\epsilon = 10^{-3}$ used in this work, this ratio is small, provided that $\tilde{L} \lesssim 20$. Furthermore, the energy loss by each BH due to pressure drag forces is roughly estimated as
\begin{equation}
\dot{E}_{\text{\tiny{drag}}} \sim  \rho_\text{\tiny env} \, m_{\text{\tiny{BH}}}^2 v^3 \sim \epsilon/\tilde{L}^{\nicefrac{9}{2}} \, .
\end{equation}
Hence, $\dot{E}_{\text{\tiny{drag}}}/\dot{E}_{\text{\tiny{GW}}} \sim \epsilon\, \tilde{L}^{\nicefrac{1}{2}}$. For $\epsilon = 10^{-3}$ and $\tilde{L} \lesssim 20$, this ratio takes similar small values. 
To summarize, matter accretion, dynamical friction, and drag forces are not expected to affect the binary dynamics after the Roche frequency.
For similar reasons, corrections arising from effects entering the waveform at negative PN orders, such as gas migration~\cite{Garg:2024oeu}, do not significantly affect the GW emission after depletion, for the values of $\tilde{L}$ we considered. 

Finally, we assess whether the assumption of zero initial spin remains valid throughout the binary evolution. While we have shown that accretion has a negligible effect on the BH masses, it could, in principle, cause significant spin-up of the binary components. However, we can demonstrate that this is not the case. Using the geodesic model for spin evolution \cite{Bardeen:1972fi, Thorne:1974ve}, the evolution of the BH spin $\chi = J/m_\BH^2$ is given by
\begin{equation}
\label{spinrate}
\dot \chi = \left( {\cal F} (\chi) - 2 \chi \right) \frac{\dot {m}_\BH}{m_\BH}\,
\end{equation}
in terms of ${\cal F} (\chi) \equiv L(m_\BH,J)/m_\BH E(m_\BH,J)$, which is only a function of $\chi$ and depends on
\begin{align}
E(m_\BH,J) &= \sqrt{1- 2 \frac{m_\BH}{3 r_\text{\tiny ISCO}}}\,, \nonumber \\
L(m_\BH,J) &= \frac{2 m_\BH }{3 \sqrt{3} } \left( 1+ 2 \sqrt{ 3 \frac{r_\text{\tiny ISCO} }{m_\BH }-2}\right)\,.
\end{align}
In these equations, the ISCO radius reads
\begin{equation}
\frac{r_\text{\tiny ISCO}}{m_\BH} = \left[ 3 + Z_2 - \sqrt{\left( 3-Z_1\right) \left( 3+Z_1+2 Z_2\right) } \right]\,, \label{ISCO}
\end{equation}
with $Z_1= 1+ \left( 1- \chi^2 \right) ^{\nicefrac{1}{3}} [ \left( 1+\chi\right)^{\nicefrac{1}{3}}+\left( 1-\chi\right)^{\nicefrac{1}{3}}]$ and $Z_2= \sqrt{3 \chi^2 + Z_1^2}$\,.

Assuming a constant accretion $\dot{m}_\BH$, we can integrate the spin evolution equation~\eqref{spinrate} to get
\begin{equation}
\int_0^{\chi_\text{\tiny{env}}} \frac{\mathrm{d} \chi}{{\cal F} (\chi) - 2 \chi}  = \frac{\dot{m}_\BH}{m_\BH} \tau \,,
\end{equation}
where $\tau$ denotes the time to coalescence from the depletion of the environment, i.e. $\tau \sim 10^{-4} f_{\text{\tiny{cut}}}^{\nicefrac{-8}{3}}m_\BH^{\nicefrac{-5}{3}} \sim 10^{-3} \,m_\BH \, \tilde{L}^4$~\cite{Peters:1963ux}, while $\chi_\text{\tiny env}$ indicates the final BH spin. 
In the small angular momentum limit (as a consequence of the mild mass growth), the integrand on the left-hand side can be simplified to
\begin{equation}
\frac{1}{{\cal F} (\chi) - 2 \chi} \approx \frac{1}{108} \left(23 \chi +12 \sqrt{6}\right)\,,
\end{equation}
such that at leading order in $\chi_\text{\tiny env}$ one gets
\begin{equation}
\chi_\text{\tiny env} \simeq 10^{-2} \left(\frac{\tilde{L}}{20}\right)^{\nicefrac{5}{2}} \left(\frac{\epsilon}{10^{-3}}\right) \,.
\end{equation}
This result shows that the spin growth is negligible for the range of parameters considered in this work, as expected from the negligible growth of the BH masses due to the suppressed accretion phase~\cite{King:1999aq}.
\section{Allowing for larger TLNs in case~1}
\label{app-B}

As discussed in Sec.~\ref{Sec:num}, calculations for case~1 are restricted to values of the tidal deformability $\tilde{\Lambda} \lesssim \mathcal{O}(20)$ to ensure that the tidal phase remains subdominant compared to the point-particle terms in the waveform, i.e.,
\begin{equation}
    \bigg|\frac{\psi_{\rm tidal}}{\psi_\text{\tiny PP}}\bigg|_{f = f_\text{\tiny ISCO}} \lesssim \mathcal{O}(1) \quad \Longleftrightarrow \quad \tilde{\Lambda} \lesssim \mathcal{O}(20)\,.
\end{equation}
This assumption is conservative since the waveform model we adopt still provides a faithful description of the signal for larger values of $\tilde{\Lambda}$ when considering frequencies below the critical value~\cite{Chia:2023tle}:
\begin{equation}
    f_{\tilde{\Lambda}} \simeq 126 \, \text{Hz} \left(\frac{20 \, \text{M}_{\odot}}{M}\right) \left(\frac{10^4}{\tilde{\Lambda}}\right)^{\nicefrac{3}{10}}\,.
    \label{fL}
\end{equation}
The relation in Eq.~\eqref{fL} is obtained by ensuring no breakdown of the stationary phase approximation for \textsc{TaylorF2} templates, which include only the inspiral evolution of the binary.

\begin{figure}[tb] 
    \includegraphics[width=\columnwidth]{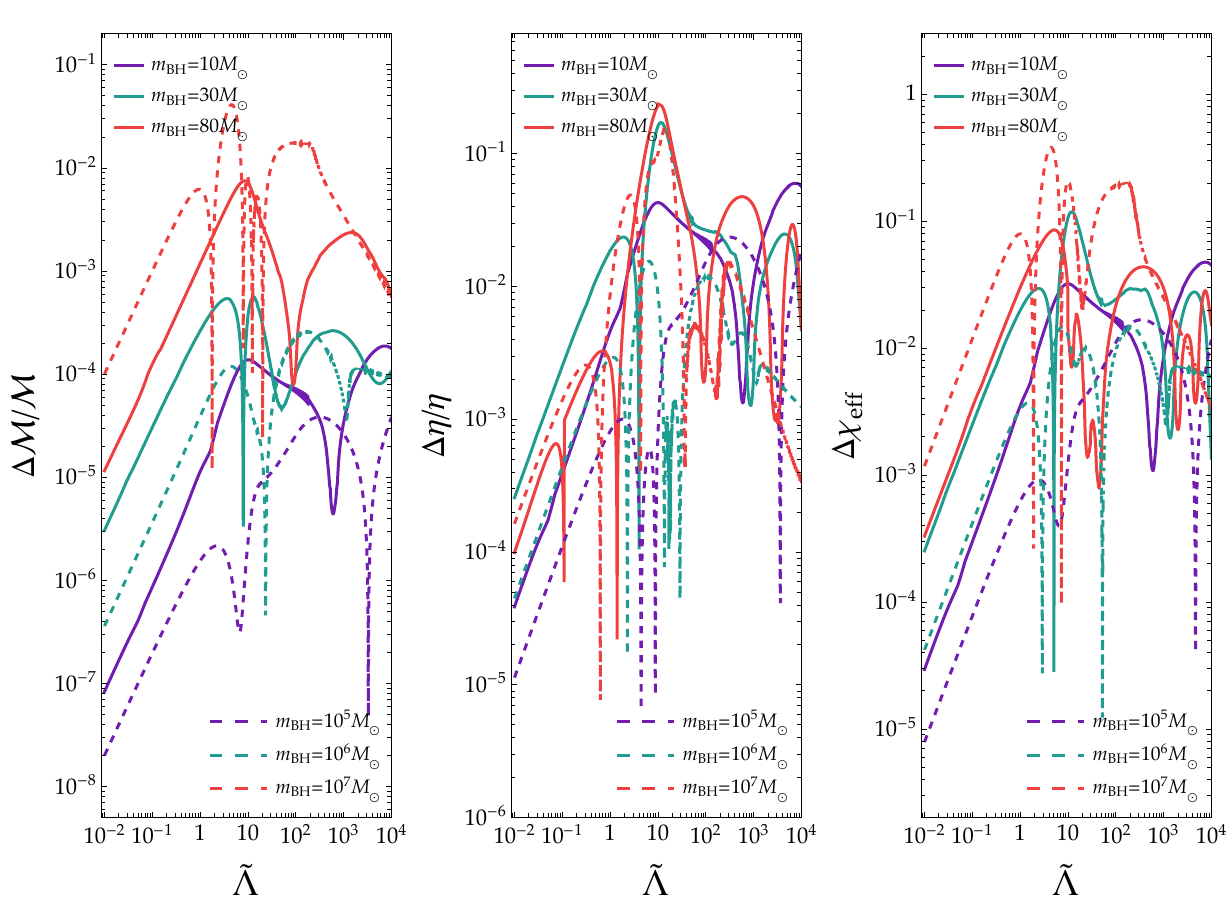}
    \caption{Same as Fig.~\ref{fig:case1} but allowing for larger values of $\tilde{\Lambda}$, following the prescription discussed in Appendix~\ref{app-B}.}
\label{fig:caseL}
\end{figure}

When restricting the analysis to the inspiral phase, $f_{\tilde{\Lambda}}$ remains above the ISCO frequency of Eq.~\eqref{eq:ISCO_freq} for $\tilde{\Lambda} \lesssim 1.5 \times 10^3$. However, for the \textsc{IMRPhenomD} waveform, which tracks the binary dynamics throughout the entire coalescence, $f_{\tilde{\Lambda}}$ becomes smaller than the maximum frequency already at values of $\tilde{\Lambda} \approx \mathcal{O}(1)$, potentially affecting the validity of the waveform to these frequency values. Equation~\eqref{fL} thus represents a natural cut-off for the waveform, provided the compactness of the inspiraling objects is close to the BH limit. More generally, one must also consider the frequency associated with contact (tidal) interactions between binary components, which depend on their compactness. For low-compactness objects, this frequency may provide an additional cut-off. 

As an exploratory investigation, we followed the prescription of Ref.~\cite{Chia:2023tle} and extended the analysis of case~1 to larger $\tilde{\Lambda}$ values than those shown in Fig.~\ref{fig:case1}, cutting the signal at frequencies $f > f_{\tilde{\Lambda}}$. The results are shown in Fig.~\ref{fig:caseL}. As expected, the bias increases linearly with the deformability for $\tilde{\Lambda} \lesssim 10^2$, potentially compromising the injected signal, particularly for heavier BHs. For larger TLNs, $\tilde{\Lambda} \gtrsim 10^2$, the amplitude of oscillations due to the mismatch between the true and recovered signals, $h_{\rm true}-h_{\rm rec} \sim (e^{i \psi_{\rm tidal}} - 1)$, increases, although the bias tends to decrease due to the cut-off imposed by $f_{\tilde{\Lambda}}$.

These results confirm that larger values of tidal deformability, when combined with a careful treatment of the integration domain, do not affect our conclusions for case~1.

However, there is an additional caveat to consider at large values of $\tilde{\Lambda}$. When the systematic error becomes comparable to or exceeds the statistical uncertainty, it is important to verify whether the following mismatch criterion,  quantifying the difference between the true and recovered waveforms, is satisfied~\cite{Lindblom:2008cm,Hughes:2018qxz}:
\begin{equation}
\label{eq:mismatch_criterion}
    1 - \underset{\{t_c, \, \phi_c\}}{\rm{max}} \, {\cal O} (h_{\rm true}, h_{\rm rec}) < \frac{N_{\rm rec}}{ 2 \rho^2 }\,,
\end{equation}
where $N_{\rm rec}$ is the number of parameters in the recovery waveform. If this condition fails, the notion of bias may become ill-defined, and such cases are better interpreted as signaling a breakdown
of the waveform model in capturing the true signal.

\begin{figure}[tbp]
    \centering
    \includegraphics[width=\columnwidth]{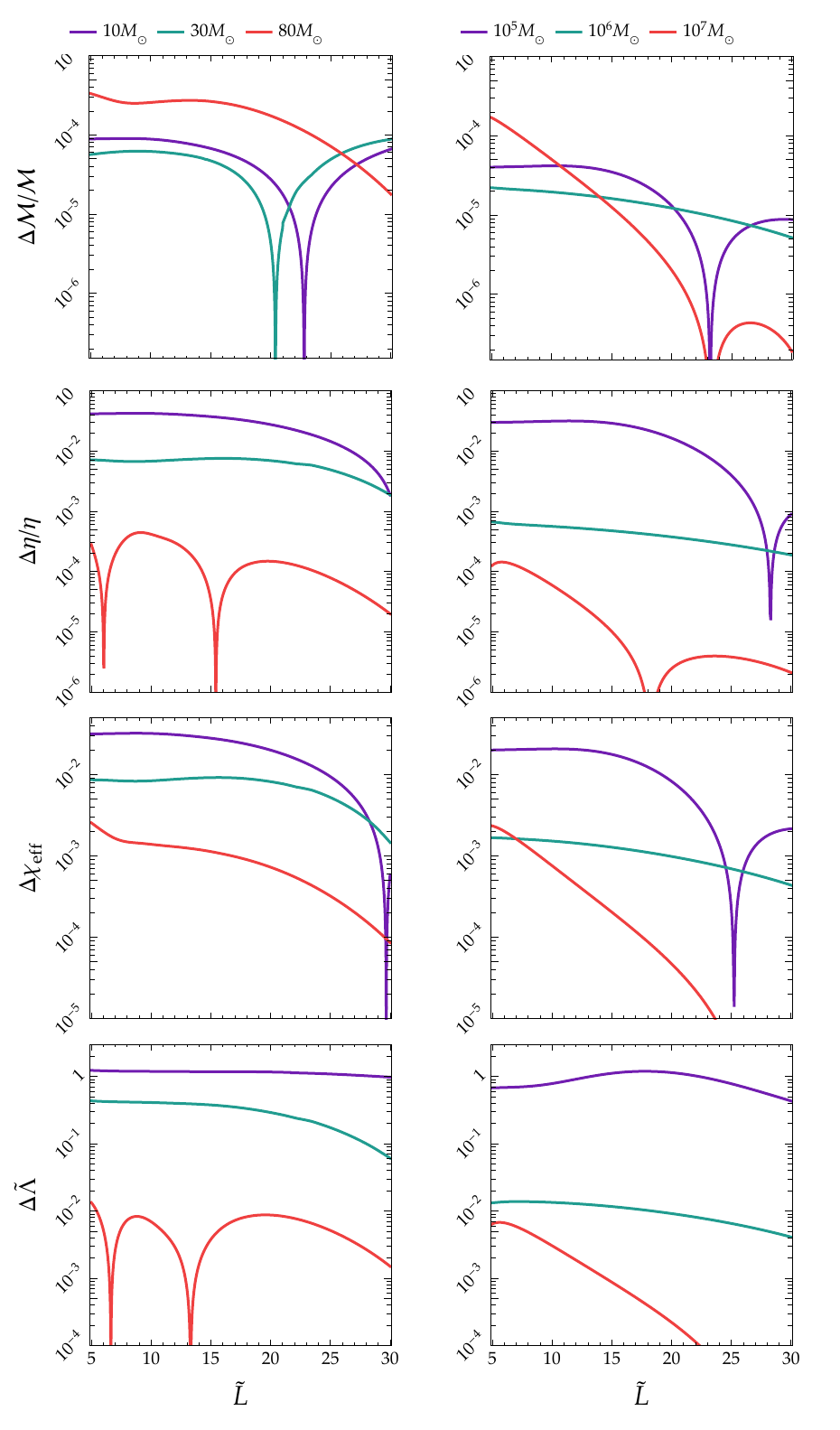}
    \caption{Same as Fig.~\ref{fig:case2}, but extending the recovery waveform parameter space by including the TLN $\tilde{\Lambda}$: see Eq.~\eqref{case4}.}
    \label{fig:case4}
\end{figure}

Equation~\eqref{eq:mismatch_criterion} translates into an upper bound on $\tilde{\Lambda}$ for a given SNR, which in the large-$\rho$ limit reads
\begin{equation}\label{eq:Lambda_bound}
    \tilde{\Lambda} \lesssim \frac{10^{-1}}{ (f_{\min} \mathcal{M})^{\nicefrac{5}{3}} \rho} \,.
\end{equation}
We verified that this condition is satisfied for the parameter space considered in the main text, though it is only marginally met for the heaviest masses ($80 \, M_\odot$ for ET and $10^7 \, M_\odot$ for LISA). In contrast, the configurations shown in Fig.~\ref{fig:caseL}, based on the extension to larger TLNs through the introduction of the critical frequency of Eq.~\eqref{fL}, do not always satisfy Eq.~\eqref{eq:Lambda_bound}, and thus the results in the large-$\tilde{\Lambda}$ limit should be interpreted with caution.

\section{Increasing parameter space in case~2}
\label{app-C}

In this appendix, we discuss an extension of case~2, in which the recovery template includes a vanishing tidal deformability within the parameters, namely
\begin{equation}\label{case4}
    \begin{aligned}
        h_{\rm rec} & = h(f,\bm{\theta})|_{\mathcal{S},\mathcal{S}_m = 1, \tilde{\Lambda} = 0}\,, \\ 
        h_{\rm true} & = h(f,\bm{\theta})\,,
    \end{aligned}
\end{equation}
where $h_{\rm true}$ describes an environment that fades during the inspiral. The biases computed from this analysis are shown in Fig.~\ref{fig:case4}. Systematic errors on $\mathcal{M},\, \eta$ and $\chi_{\rm eff}$ show a trend similar to case~2 (cf. Fig.~\ref{fig:case2}). The small differences can be attributed to correlations with the tidal deformability, which is now part of the waveform parameters, and affect the Fisher matrix and the bias.

In the bottom row of Fig.~\ref{fig:case4}, we also present the bias on $\tilde{\Lambda}$, which is absent in the analysis of case~2, where the tidal deformability is not included as a waveform parameter. The bias on $\tilde{\Lambda}$ can be significant for the lightest BHs, as the cut-off frequency occurs at higher values, allowing the mismatch to accumulate over a larger number of cycles. For the same reference systems at $d_L=1\,{\rm Gpc}$ used as examples in the text, we find that the systematic uncertainty on $\tilde{\Lambda}$ can exceed the statistical one by as much as an order of magnitude for the lighter ET and LISA sources at small $\tilde{L}$. This is opposite to the trend found in case~3, where the larger masses show higher systematic biases as compared to the statistical uncertainties. The difference can be explained as follows. In case~3 the recovery waveform assumes a constant environment neglecting the fading, which impacts the signal \emph{after} $f_\text{\tiny cut}$, while in the extension of case~2 discussed here the opposite is true. Thus, in this latter case, the longer signals (i.e. the lower-mass binaries) have more cycles to accumulate mismatch with respect to the true waveform, resulting in larger biases.

\bibliography{TLNsystematics}

\end{document}